\documentclass[twocolumn]{jpsj3}
\usepackage{multicol}
\usepackage{times}
\usepackage{color}
\usepackage{bm}

\title{
Spin-orbit Coupling and Multiple phases in Spin-triplet Superconductor Sr$_2$RuO$_4$ 
}

\author{Youichi Yanase$^{1,2}$\thanks{E-mail address: yanase@phys.sc.niigata-u.ac.jp}, 
Shuhei Takamatsu$^{2}$, and Masafumi Udagawa$^{3}$ 
}
\inst{
$^{1}$Department of Physics, Niigata University, 8050 Ikarashi, Nishi-ku, Niigata 950-2181, Japan
\\
$^{2}$Garaduate School of Science and Technology, Niigata University, 8050 Ikarashi, Nishi-ku, Niigata 950-2181, Japan
\\
$^{3}$Department of Applied Physics, University of Tokyo, 7-3-1 Hngo, Bunkyo-ku, Tokyo 113-8656, Japan
}
\abst{
We study the spin-orbit coupling in spin-triplet Cooper pairs and clarify multiple superconducting (SC) phases 
in Sr$_2$RuO$_4$.  
Based on the analysis of the three-orbital Hubbard model with atomic LS coupling, we show some 
selection rules of the spin-orbit coupling in Cooper pairs. 
The spin-orbit coupling is small when the two-dimensional $\gamma$-band is the main cause of the superconductivity, 
although the LS coupling is much larger than the SC gap. 
Considering this case, we investigate multiple SC transitions in the magnetic fields for both 
$H \parallel [001]$ and $H \parallel [100]$ using the Ginzburg-Landau theory and the quasi-classical theory. 
Rich phase diagrams are obtained because the spin degree of freedom in Cooper pairs is not quenched 
by the spin-orbit coupling. Experimental indications for the multiple phases in Sr$_2$RuO$_4$ are discussed. 
}

\kword{spin-triplet superconductivity, spin-orbit coupling, multiple phases, Sr$_2$RuO$_4$}

\begin{document}
\maketitle

\renewcommand{\k}{{\bf k}}
\renewcommand{\r}{{\bf r}}
\newcommand{\dd}{{\bf d}}
\newcommand{\kk}{{\bf k'}}
\newcommand{\kkk}{{\bf k''}}
\newcommand{\q}{{\bf q}}
\newcommand{\Q}{{\bf Q}}
\newcommand{\e}{\varepsilon}
\newcommand{\ee}{e}
\newcommand{\s}{{\mit{\it \Sigma}}}
\newcommand{\J}{\mbox{\boldmath$J$}}
\newcommand{\vv}{\mbox{\boldmath$v$}}
\newcommand{\Jh}{J}
\newcommand{\LL}{\mbox{\boldmath$L$}}
\renewcommand{\SS}{\mbox{\boldmath$S$}}
\newcommand{\MM}{\mbox{\boldmath$M$}}
\newcommand{\g}{\mbox{\boldmath$g$}}
\newcommand{\Tc}{$T_{\rm c}$ }
\newcommand{\Tcf}{$T_{\rm c}$}
\newcommand{\Hc}{$H_{\rm c2}$ }
\newcommand{\Hcf}{$H_{\rm c2}$}
\newcommand{\etal}{{\it et al.}: }
\newcommand{\SRO}{Sr$_2$RuO$_4$ }
\newcommand{\SROf}{Sr$_2$RuO$_4$}
\newcommand{\px}{p_{x}}
\newcommand{\py}{p_{y}}

\section{Introduction}

In this short review, we study the spin-orbit coupling and multiple phases in spin-triplet superconductors. 
The spin-orbit coupling plays a crucial role in determining the spin-triplet pairing state.
Discussions are particularly focused on Sr$_2$RuO$_4$~\cite{Maeno1994} which is an  
established candidate of the spin-triplet superconductor~\cite{Mackenzie2003,Maeno2012}  
in addition to the superfluid $^{3}$He~\cite{Leggett1975}, heavy fermion superconductor 
UPt$_{3}$~\cite{Sauls,Joynt,Tou}, and ferromagnetic superconductors, UGe$_2$, URhGe, 
and UCoGe~\cite{Aoki}. 

Because of the simple electronic structure of Sr$_2$RuO$_4$ compared with the U-based 
heavy fermion superconductors, studies on Sr$_2$RuO$_4$ for these two decades 
have made noticeable progress in the microscopic understanding of spin-triplet superconductivity. 
Low energy quasiparticles in Sr$_2$RuO$_4$ are described by the two-dimensional 
tight-binding model for the three $t_{\rm 2g}$-orbitals in Ru ions on the tetragonal crystal 
with $D_{4h}$ point group symmetry. 
Indeed, electronic and SC properties of Sr$_2$RuO$_4$ have been  
elucidated on the basis of the three-orbital Hubbard model~\cite{Maeno2012}. 
Those microscopic theories provided several clear understandings of spin-triplet superconductivity 
which have not been obtained in the studies of f-electron systems~\cite{Sauls,Joynt}.

In the first part of this article (\S3), we elucidate how the spin-orbit coupling 
in spin-triplet Cooper pairs arises from the atomic LS coupling of electrons. 
Analysis of the three-orbital Hubbard model shows that the spin-orbit coupling in Cooper pairs is small, 
in spite of the large LS coupling compared with the energy scale of superconductivity. 
We also demonstrate some selection rules which derive from the symmetry of local 
electron orbital. The selection rules sometimes determine the $d$-vector, namely, 
the order parameter of spin-triplet superconductivity. 
In the second part (\S4 and \S5), we study the multiple SC phases 
in the magnetic field on the basis of the Ginzburg-Landau (GL) theory and quasi-classical theory 
derived from the three-orbital Hubbard model. 
When the spin-orbit coupling is small but finite, multiple SC phases appear
in the magnetic-field-temperature ($H$-$T$) phase diagram. 
We clarify the pairing state in \SRO and discuss the experimental results. 
Several indications for the multiple phases as well as some unresolved issues are discussed.

\section{Spin-triplet Superconductivity in Tetragonal Crystals} 
 
First, we review the general aspects of spin-triplet superconductivity 
and define the ``spin-orbit coupling in Cooper pairs''. 
Since the Cooper pairs have total spin $S=1$, the order parameter of spin-triplet superconductors 
is described by the three component vector, 
$\dd =(d_{x},d_{y},d_{z})$~\cite{Leggett1975,Sigrist-Ueda}, 
\begin{equation}
\left(
\begin{array}{cc}
\Delta_{\uparrow\uparrow} & \Delta_{\uparrow\downarrow}
\\
\Delta_{\downarrow\uparrow} & \Delta_{\downarrow\downarrow}
\end{array}
\right)
= 
\left(
\begin{array}{cc}
- d_{x} + i d_{y} & d_{z} 
\\
d_{z} & d_{x} + i d_{y}
\end{array}
\right). 
\label{eq:d-vector}
\end{equation}
The $p$-wave superconductivity in the tetragonal crystal 
also has two orbital components, that is $\px$ and $\py$, and therefore, 
the SC state is represented by the $2 \times 3 = 6$ component order parameters. 
In the presence of the spin-orbit coupling, the spin is entangled with the orbital, and 
the SC states are classified on the basis of the point group~\cite{Sigrist-Ueda}. 
For the $D_{4h}$ point group symmetry, the SC state belongs to the two-dimensional irreducible representation 
$E_{u}$, or to four one-dimensional representations, $A_{1u}$,  $A_{2u}$, $B_{1u}$, and $B_{2u}$, 
as summarized in Table I.
Some experiments of Sr$_2$RuO$_4$, such as $\mu$SR~\cite{Luke1998} and Kerr rotation~\cite{Xia2006}, 
observe the spontaneous time-reversal symmetry breaking (TRSB) in the SC 
state~\cite{Mackenzie2003,Maeno2012} and indicate the SC state belonging to the $E_{u}$ representation. 
This means that the spin-orbit coupling favors the ``chiral SC state'', 
namely $\dd = (\px \pm i\py) \hat{z}$. On the other hand, the other SC states belonging to 
$A_{1u}$,  $A_{2u}$, $B_{1u}$, or $B_{2u}$ representation are called ``helical SC state''.

We would like to stress that the spin-orbit coupling particularly plays a crucial role 
in the tetragonal crystal, in contrast to the cubic crystals and the rotationally-symmetric superfluid $^{3}$He. 
The B-phase of superfluid $^{3}$He is stable at low temperatures even in the absence of the 
spin-orbit coupling so that the condensation energy is maximized through the isotropic  
excitation gap.~\cite{Leggett1975} 
Then, the weak spin-orbit coupling arising from the dipole interaction plays a minor role. 
On the other hand, the spin-orbit coupling plays an essential role in determining the SC state 
of Sr$_2$RuO$_4$ even when the spin-orbit coupling is small, because 
the condensation energy is equivalent between the SC states in Table I in the absence of the spin-orbit coupling.
Therefore, it is crucial to investigate the spin-orbit coupling in Cooper pairs for the study of 
spin-triplet SC state in Sr$_2$RuO$_4$. 
This is also the case in the other spin-triplet superconductors except for those in the cubic crystals. 

\begin{table}[htbp]
\begin{center}
{\renewcommand\arraystretch{1.5}
\begin{tabular}{cccc}
Irreducible representation & Order parameter & Dimension  
\\
\hline
 $A_{1u}$ & $\dd = \px \hat{x} + \py \hat{y}$ & 1 
\\
\hline
 $B_{1u}$ & $\dd = \px \hat{x} - \py \hat{y}$ & 1 
\\
\hline
$A_{2u}$ & $\dd = \py \hat{x} - \px \hat{y}$ & 1 
\\
\hline
$B_{2u}$ & $\dd = \py \hat{x} + \px \hat{y}$ & 1 
\\
\hline
$E_{u}$ & $\dd = (\px \pm i\py) \hat{z}$ & 2  
\\
\hline
\end{tabular}
}
\end{center}
\caption{
Spin-triplet SC states in tetragonal crystals with $D_{4h}$ point group symmetry. 
We show the order parameters belonging to the irreducible representation, 
$A_{1u}$,  $A_{2u}$, $B_{1u}$, $B_{2u}$, and $E_{u}$. Their dimension is also shown. 
}
\label{table1}
\end{table}
For the aim of a coherent discussion, 
we here define the spin-orbit coupling in Cooper pairs. That is denoted as 
$\eta^{\Gamma \Gamma'} = (T_{\rm c}^{\Gamma} - T_{\rm c}^{\Gamma'})/T_{\rm c}$, where $\Gamma$ and $\Gamma'$ label the 
irreducible representation. Thus, the spin-orbit coupling in Cooper pairs represents the difference 
of transition temperature between irreducible representations. Clearly, $\eta^{\Gamma \Gamma'} =0$ when 
the spin $SU(2)$ symmetry is conserved. On the other hand, the violation of $SU(2)$ symmetry 
gives rise to a finite spin-orbit coupling, $\eta^{\Gamma \Gamma'}$. 

When the transition temperature is highest for the pairing state belonging to an 
irreducible representation $\Gamma_{0}$, such SC state is stabilized below 
$T_{\rm c} = T_{\rm c}^{\Gamma_0}$. 
Among four independent spin-orbit couplings in the $D_{4h}$ point group symmetry, 
the most important one is $\eta = \eta^{\Gamma_0 \Gamma_1}$ where $\Gamma_1$ is the irreducible representation 
having the second highest transition temperature. The magnitude of $\eta$ represents the 
anisotropy of Cooper pairs in the spin space. 
Generally speaking, multiple SC phases appear in the $H$-$T$ phase diagram 
when the ``anisotropy'' $\eta$ is small. 
Other spin-orbit couplings also play important roles in determining the multiple phases (see \S4).

\section{Spin-orbit Coupling in Spin-triplet Cooper Pairs}

Next, we discuss the microscopic aspects of the spin-orbit coupling in Cooper pairs. 
Although we see some similarities between the superfluid $^3$He and spin-triplet superconductors, 
the origin and properties of the spin-orbit coupling are quite different between them. 
It has been clarified that the relevant spin-orbit coupling in $^3$He is the dipole 
interaction.~\cite{Leggett1975} On the other hand, electrons in the crystals are affected 
by the so-called LS coupling which originates from the relativistic motion of electrons near nuclei~\cite{Harima}. 
It has been established that the spin anisotropy of electrons mainly originates from the LS coupling 
in the solid state physics~\cite{Yosida_book}. 
Therefore, it is reasonable that the LS coupling gives rise to the leading spin-orbit coupling 
in spin-triplet Cooper pairs. 
However, the relation between the LS coupling and the spin-orbit coupling in Cooper pairs is  non-trivial. 
In this section, we clarify how the spin-orbit coupling in Cooper pairs arises from 
the LS coupling of electrons~\cite{Yanase2003,Yanase2005}.

\subsection{Three-orbital Hubbard model}

Our discussions are based on the theoretical analysis of two-dimensional three-orbital 
Hubbard model which reproduces the band structure of Sr$_2$RuO$_4$~\cite{Mackenzie2003,Maeno2012,Oguchi,Singh}. 
We here focus Sr$_2$RuO$_4$ as a typical example, however, the following results on the spin-orbit coupling 
in Cooper pairs are valid for other 3d and 4d electron systems too. The model is 
\begin{eqnarray}
\label{t2g}
&& \hspace*{-10mm}  
H = 
H_{\rm kin} + H_{\rm hyb} + H_{\rm CEF} + H_{\rm LS} + H_{\rm I}, 
\end{eqnarray}
where 
$
H_{\rm kin} = 
     \sum_{\k} \sum_{m=1,2,3} \sum_{s = \uparrow,\downarrow} 
\e_{m}(\k) c_{\k m s}^{\dag}c_{\k m s} 
$
is the kinetic energy, 
$
H_{\rm hyb} = \sum_{\k} \sum_{s} 
[V(\k) c_{\k{\rm 1}s}^{\dag} c_{\k{\rm 2}s} + {\rm h.c.}]  
$
describes the intersite hybridization between the d$_{\rm yz}$- and 
d$_{\rm zx}$-orbitals, 
$
H_{\rm CEF} = \Delta \sum_{\k}  \sum_{s} c_{\k 3 s}^{\dag} c_{\k 3 s} 
$
is the crystal electric field term, 
and 
$
H_{\rm LS} = \lambda \sum_{i} \LL_i \cdot \SS_i 
$
represents the LS coupling. 
The d$_{\rm yz}$-, d$_{\rm zx}$-, and d$_{\rm xy}$-orbitals are denoted 
by the indices $m=$ 1, 2, and 3, respectively. 
We will show that the absence of intersite hybridization between 
d$_{\rm xy}$- and d$_{\rm yz}$/d$_{\rm zx}$-orbitals plays an important role 
in the spin-orbit coupling of Cooper pairs. 
Taking account of the symmetry of $t_{2g}$-orbitals, we adopt the tight-binding form, 
$\e_{1}(\k)  = -2 t_4 \cos k_{\rm x} -2 t_3 \cos k_{\rm y}$, 
$\e_{2}(\k)  = -2 t_3 \cos k_{\rm x} -2 t_4 \cos k_{\rm y}$, 
$\e_{3}(\k)  = -2 t_1 (\cos k_{\rm x} + \cos k_{\rm y}) -4 t_2 \cos k_{\rm x} \cos k_{\rm y}$, 
and 
$V(\k) = 4 t_5 \sin k_{\rm x} \sin k_{\rm y}$.

%
\begin{figure}[htbp]
\begin{center}
\includegraphics[width=6.5cm]{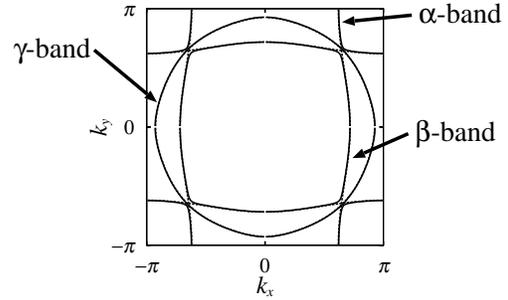}
\end{center}
\caption{Fermi surfaces of the $\alpha$-, $\beta$-, and $\gamma$-bands in the three-orbital Hubbard model. 
The dashed lines show the Fermi surfaces in the absence of the LS coupling $H_{\rm LS}$ and 
intersite hybridization term $H_{\rm hyb}$. 
The band gap opens along $\k \parallel [110]$ owing to $H_{\rm LS}$ and $H_{\rm hyb}$  (solid lines). 
Tight-binding parameters are shown in Ref.~\citen{Yanase2003}. 
These Fermi surfaces reproduce the cylindrical Fermi surfaces of 
Sr$_2$RuO$_4$~\cite{Mackenzie2003,Maeno2012,Oguchi,Singh}. 
}
\label{fig1}
\end{figure}
This model appropriately reproduces the band structure of 
Sr$_2$RuO$_4$ which has been elucidated by the first principle band structure 
calculation~\cite{Oguchi,Singh} as well as by the de Haas-van Alphen oscillation measurements
and angle-resolved photo-emission spectroscopy.~\cite{Macken zie2003,Maeno2012}  
As shown in Fig.~1, 
we see the quasi-two-dimensional Fermi surface of the d$_{\rm xy}$-orbital ($\gamma$-band) 
and two quasi-one-dimensional Fermi surfaces consisting of the (d$_{\rm yz}$, d$_{\rm zx}$)-orbitals 
($\alpha$- and $\beta$-bands). The electron correlation effect renormalizes the band 
structure,~\cite{Mackenzie2003,Maeno2012} but hardly changes the Fermi surfaces.

Although the BCS theory assumed the $s$-wave superconductivity induced by the electron-phonon coupling, 
the unconventional non-$s$-wave superconductivity occurs through the Coulomb interactions 
in the strongly correlated electron systems~\cite{Yanase_review}. 
Thus, we take into account the on-site Coulomb interaction term $H_{\rm I}$ which consists of 
the intraorbital repulsion $U$, interorbital repulsion $U'$, Hund's rule coupling $J$, and pair hopping $J'$. 
Two spin-triplet SC states have been obtained by the theoretical analysis of the 
three-orbital Hubbard model. One is the $p$-wave SC state which is mainly caused 
by the quasi-two-dimensional $\gamma$-band~\cite{Nomura2002-2,Hoshihara,Wang}. 
The other is the $p$- or $f$-wave state mainly due to the quasi-one-dimensional 
($\alpha, \beta$)-bands~\cite{Takimoto,Raghu}.  
The partial density of states (DOS) of the bands determines which SC state is stable. 
Indeed, our calculation showed the crossover from the quasi-two-dimensional superconductivity to the 
quasi-one-dimensional superconductivity by tuning the tight-binding parameters so as to 
decrease the partial DOS of $\gamma$-band~\cite{Yanase2003}. 
Thus, the superconductivity is mainly induced by the ``active orbital'' 
as proposed by Agterberg {\it et al.}~\cite{Agterberg1997} 
When we choose the realistic parameters so as to reproduce the 57\% partial DOS 
in the $\gamma$-band,~\cite{Mackenzie2003,Maeno2012} 
both perturbation theory~\cite{Nomura2002-2,Yanase2003} and functional renormalization group 
theory~\cite{Wang} support the superconductivity driven by the $\gamma$ Fermi surface. 
In all cases, the spin-triplet SC states summarized in Table I are degenerate in the 
absence of the LS coupling, because the spin $SU(2)$ symmetry is conserved.

\subsection{Order estimation of spin-orbit coupling}

Now we move on to the role of LS coupling, which is the main topic of this article. 
Although it is not difficult to non-perturbatively deal with the LS coupling term, we here adopt 
the perturbation expansion for $\lambda$ by which we obtain some selection rules in the following way.  
The discussion is based on the hierarchy of energy scales in the 3d and 4d electron systems, 
\begin{eqnarray}
\label{hierarcy}
&& \hspace*{-2mm}  
T_{\rm c} \ll \lambda \ll E_{\rm F}. 
\end{eqnarray}
In the case of Sr$_2$RuO$_4$, the LS coupling $\lambda \sim 100$ K is much larger than the 
transition temperature of superconductivity $T_{\rm c} \sim 1$ K, but much smaller than the 
Fermi energy $E_{\rm F} = 1000 \sim 10000$ K. 
Now let us consider the perturbation expansion of the spin-orbit coupling in Cooper pairs, 
\begin{eqnarray}
\label{LS_expansion}
&& \hspace*{-2mm}  
\eta = \sum_{n=1}^{\infty} \sum_{m=1}^{\infty} A_{nm} (\lambda/T_{\rm c})^{n} (\lambda/E_{\rm F})^{m}. 
\end{eqnarray}
When the coefficients $A_{nm} \,\, (n \ge 1)$ are finite,
this expansion is unreliable because the expansion parameter $\lambda/T_{\rm c}$ is huge. 
However, we find $A_{nm} =0$ for $n \ge 1$.~\cite{Yanase2003,Yanase2005} 
This property is guaranteed by the inversion symmetry of the system as we discuss in \S3.5. 
Thus, we obtain the quantitatively reliable perturbation expansion of $\eta$ 
for the small parameter $\lambda/E_{\rm F} \ll 1$, as $\eta = \sum_{m=1}^{\infty} A_{0m} (\lambda/E_{\rm F})^{m}$. 
At the same time we understand that the spin-orbit coupling in Cooper pairs $\eta$ 
is small when $\lambda/E_{\rm F} \ll 1$, even though the LS coupling is much larger than 
the energy scale of superconductivity.

\subsection{Selection rules}

Next, we show the selection rules of the spin-orbit coupling in Cooper pairs. 
We here discuss the following two SC phases in a separate way;  
(1) superconductivity in the quasi-one-dimensional ($\alpha, \beta$)-bands 
and (2) that in the quasi-two-dimensional $\gamma$-band. 
Indeed, all bands are superconducting owing to the inter-band proximity effect, and therefore, 
the spin-orbit coupling in Cooper pairs are obtained by adding the contributions of $\alpha$, $\beta$, and 
$\gamma$ bands. 
However, the SC properties are mainly determined by the active band having a large SC gap, 
since the orbital dependent SC phases~\cite{Agterberg1997} are likely stabilized in Sr$_2$RuO$_4$ (see \S3.1).

\begin{figure}[htbp]
\begin{center}
\includegraphics[width=8.0cm]{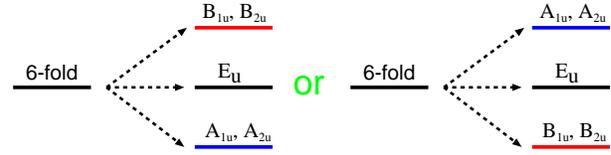}
\end{center}
\caption{(Color online) 
Illustration of the selection rule for the SC state driven by the (d$_{\rm yz}$, d$_{\rm zx}$)-orbitals. 
We show the energy levels of 6 spin-triplet SC states. The 6-fold degeneracy is lifted by the first order term 
of LS coupling. 
One of the doublet, ($A_{1u}$, $A_{2u}$) or ($B_{1u}$, $B_{2u}$), has the lowest energy. The 2-fold degeneracy 
in the doublet is lifted by the higher order terms. 
}
\label{fig2}
\end{figure}

We begin with the discussion of the case (1). In this case, the leading order terms of 
the spin-orbit coupling $\eta^{\Gamma\Gamma'}$ are first order in $\lambda/E_{\rm F}$. 
Analyzing the Eliashberg equation~\cite{Yanase_review} for the three-orbital Hubbard model, 
it is shown that the first order terms of the irreducible vertex in the particle-particle channel 
have the particular symmetry~\cite{Yanase2003,Yanase2005}. Those terms have the $d_{\rm xy}$ symmetry 
in the momentum space, and conserves the $z$-component of the total spin. The selection rule which is 
schematically shown in Fig.~2 arises from this symmetry. 
The first order terms do not lift the degeneracy between the $A_{1u}$ and $A_{2u}$ states 
and between the $B_{1u}$ and $B_{2u}$ states. On the other hand, the degeneracy between the two doublet is lifted,
and one of the doublet has the lowest energy. 
This selection rule is explicitly described as $\eta^{A_{1u}E_{u}} = \eta^{A_{2u}E_{u}} = 
- \eta^{B_{1u}E_{u}} = - \eta^{B_{2u}E_{u}} = O(\lambda/E_{\rm F})$. 
The degeneracy of $A_{1u}$ and $A_{2u}$ states ($B_{1u}$ and $B_{2u}$ states) is 
slightly lifted by the second order term, as $\eta^{A_{1u}A_{2u}} = O(\lambda^2/E_{\rm F}^2)$ 
($\eta^{B_{1u}B_{2u}} = O(\lambda^2/E_{\rm F}^2)$). 
Although the signs of $\eta^{A_{1u}E_{u}}$ and $\eta^{A_{1u}A_{2u}}$ depend on the electronic structure, 
we obtain  an exact conclusion; One of the helical SC states is stabilized by the 
LS coupling. In other words, the chiral SC state belonging to the $E_{u}$ representation 
can not be stable when the ($\alpha, \beta$)-bands are mainly superconducting. 
This feature was also pointed out by the semi-microscopic calculation~\cite{Ng-Sigrist}.

Importantly, these selection rules are independent of the Coulomb interactions. 
Indeed, we confirmed that the selection rules for $\eta^{\Gamma\Gamma'}$ are satisfied in all order of 
perturbation terms for Coulomb interactions $U$, $U'$, $\Jh$, and $J'$~\cite{Yanase2003,Yanase2005}. 
This feature is not altered even when we take into account the long-range Coulomb interaction. 
Thus, the leading order term of spin-orbit coupling in Cooper pairs obeys the selection rule 
which is independent of the electron correlation. 
This is in sharp contrast to the fact that the pairing interaction leading to the unconventional 
superconductivity depends on the band structures, Coulomb interactions, and so on~\cite{Yanase_review}.
This means that the spin-orbit coupling in Cooper pairs is not closely related 
with the mechanism of Cooper pairing.

\begin{figure}[htbp]
\begin{center}
\includegraphics[width=8.5cm]{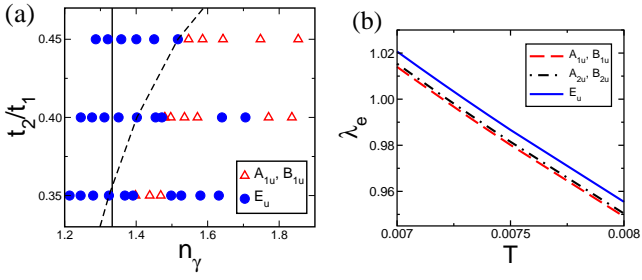}
\end{center}
\caption{(Color online) 
(a) Phase diagram of the three-orbital Hubbard model~\cite{Yanase2003} 
for a tight-binding parameter $t_2/t_1$ and the electron density in the $\gamma$-band, $n_{\gamma}$. 
The other parameters are chosen to be $(t_1, t_3, t_4, t_5, \lambda, \Delta, U, U', J, J') 
= (1, 1.25, 0.1, 0, 0.2, -0.3, 5, 1.5, 1, 1)$. 
Circles show the $E_{u}$ state, and triangles show the ($A_{1u}, B_{1u}$) state. 
(b) Temperature dependence of eigenvalues of Eliashberg equation for the 
($A_{1u}, B_{1u}$) state (dashed line), ($A_{2u}, B_{2u}$) state (dot-dashed line), 
and $E_{u}$ state (solid line), respectively. We assume $t_2/t_1 =0.4 $ and $n_{\gamma} = 1.37$.  
}
\label{fig3}
\end{figure}

We turn to the discussion of the case (2). When the superconductivity is mainly caused by the 
quasi-two-dimensional $\gamma$-band, the first order term with respect to the LS coupling vanishes. 
We find this feature by analyzing the Eliashberg equation. 
The first order terms of the irreducible vertex lead to the inter-band Cooper pairing, and they are 
negligible when $T_{\rm c} \ll E_{\rm F}$. 
This is also the selection rule and comes form the fact that the intersite hybridization between 
d$_{\rm xy}$- and d$_{\rm yz}$/d$_{\rm zx}$-orbitals vanishes owing to the mirror symmetry along the {\it c}-axis. 
Although such hybridization terms appear in the three-dimensional model, 
they do not alter the selection rule.

As the leading order term is roughly estimated as 
$\eta^{\Gamma\Gamma'} = O(\lambda^2/E_{\rm F}^2) \sim 0.01 $ for $\lambda/E_{\rm F} \sim 0.1$, 
the spin-orbit couplings in Cooper pairs $\eta^{\Gamma\Gamma'}$ are small when the $\gamma$-band 
is mainly superconducting. In order to investigate this small spin-orbit coupling, we are required to 
solve the three-orbital Hubbard model with use of some approximate treatments of Coulomb interactions.  
We do not find any selection rule for the second order terms in $\lambda$ 
except for the accidental degeneracy between the $A_{1u}$ and $B_{1u}$ states and 
between the $A_{2u}$ and $B_{2u}$ states. 
Using the perturbation theory for Coulomb interactions up to the third order, 
we solved the linearized Eliashberg equation and obtained the results in Fig.~3~\cite{Yanase2003}. 
Figure~3(a) shows the phase diagram against the tight-binding parameter $t_2/t_1$ and 
the number density of electrons in the $\gamma$-band, $n_{\gamma}$. 
It is shown that the spin-triplet SC state 
depends on these relevant parameters. For realistic parameters of Sr$_2$RuO$_4$, 
namely $n_{\gamma} \sim 1.33$ and $t_2/t_1 \sim 0.4$, the $E_{u}$ state is stable, although 
the $A_{1u}$ or $B_{1u}$ state is stabilized in a part of the phase diagram.

Figure~3(b) shows the temperature dependence of the eigenvalue of the linearized Eliashberg equation 
$\lambda_{\rm e}$ for the $E_{u}$, ($A_{1u}$, $B_{1u}$), and ($A_{2u}$, $B_{2u}$) states. 
The \Tc of each superconducting state is obtained by the criterion, $\lambda_{\rm e}=1$. 
We see that the $E_{u}$ state has the highest \Tcf, 
but the splitting of $T_{\rm c}$ is small $\eta^{E_{u}A_{1u}} = 0.013$. 
Thus, the spin-orbit coupling in Cooper pairs is small as we expected from the order estimation, 
although the LS coupling ($\lambda =0.2$) is much larger than the transition temperature 
of superconductivity ($T_{\rm c} = 0.0073$).

As we have shown above, when the spin-triplet SC state is induced by the $\gamma$-band  
``a small spin-orbit coupling in Cooper pairs favors the chiral SC state 
$\dd = (p_{x} \pm i p_{y}) \hat{z}$ ($E_{u}$ state)''.  
The same result was obtained by the recent calculation based on the functional renormalization 
group theory~\cite{Wang}. 
We show this result in the Table~II, although it is not obtained by the selection rule. 
On the other hand, it has been shown that the helical SC state 
is stabilized when the Coulomb interaction on Oxygen ions is large~\cite{Yoshioka}. 
It is reasonable that the pairing state depends on the electron interaction, 
because we do not find any selection rule in this case.

Table~II summarizes the d-vector of spin-triplet Cooper pairs which we obtained~\cite{Yanase2005}. 
We also show the case of the hexagonal crystal with $D_{6h}$ point group symmetry. 
It is shown that the anisotropy $\eta$ and the direction of $d$-vector are determined by the symmetries of 
crystal lattice, local electron orbital, and superconductivity. 
Interestingly, we find a similarity between the tetragonal crystal and hexagonal crystal. 
When the SC is induced by the $A_{1g}$-orbital in the latter, the $\eta$ is in the second order of $\lambda/E_{\rm F}$ 
as in the case of d$_{xy}$-orbital in the former.  
On the other hand, the first order term in $\lambda/E_{\rm F}$ stabilizes the d-vector parallel to the {\it ab}-plane 
when the $p$-wave SC occurs in the $E_{g}$-orbitals of the hexagonal crystal. 
What is different from the tetragonal lattice appears in the last column of Table~II. 
In contrast to the tetragonal crystal, the $f$-wave superconductivity is distinguished from the $p$-wave 
superconductivity in the hexagonal crystal.  
In the $f$-wave SC state, the first order term in $\lambda/E_{\rm F}$ vanishes, and we can not determine the 
d-vector by the selection rule. 
Thus, the orbital symmetry of superconductivity also plays an important role for the 
spin-orbit coupling in Cooper pairs.

\subsection{Spin-orbit coupling in Sr$_2$RuO$_4$}

We discuss the experimental results indicating the SC state of Sr$_2$RuO$_4$. 
First, the spontaneous TRSB observed in the $\mu$SR~\cite{Luke1998} and Kerr rotation~\cite{Xia2006} 
measurements implies that the $E_{u}$ state is stabilized at zero magnetic field. 
This finding is compatible with our results on the superconductivity in 
the quasi-two-dimensional $\gamma$-band~\cite{Yanase2003,Wang}. On the other hand, the TRSB is 
incompatible with the selection rule for the quasi-one-dimensional ($\alpha, \beta$)-bands, which 
does not allow the $E_{u}$ state to be stabilized at zero magnetic field. 

Although the interpretation of the $\mu$SR~\cite{Luke1998} and Kerr rotation~\cite{Xia2006} 
data are still under the discussion~\cite{Kallin2012}, 
the magnitude of the spin-orbit coupling is also consistent with the superconductivity in the 
quasi-two-dimensional $\gamma$-band.  
A small spin-orbit coupling below $\eta < 0.01$ is indicated by several experiments. 
The nuclear magnetic resonance (NMR) measurements have shown the temperature-independent Knight shift 
through the SC transition temperature in both magnetic field directions 
along the {\it ab}-plane and along the {\it c}-axis~\cite{Ishida,Murakawa2004,Murakawa2007}. 
This observation shows that 
the spin-orbit coupling in Cooper pairs is so small that the $d$-vector rotates in the magnetic field. 
The magnitude of spin-orbit coupling is estimated to be $\eta \sim 0.001$ 
according to the temperature independent Knight shift data at $H^{c} = 0.02$T~\cite{Murakawa2004}. 
Such a tiny spin-orbit coupling is not incompatible with our calculation for 
the superconductivity in the $\gamma$-band. 
We obtained $\eta = 0.01$ for the LS coupling $\lambda = 50$ meV in \S3.4, but the LS coupling may be 
smaller, because the LS coupling of Sr$_2$RuO$_4$ is reduced by the strong hybridization of Ru and O ions 
as demonstrated by another NMR measurement~\cite{Kitagawa2007}. 
Furthermore, the spin-orbit coupling $\eta$ decreases because of the competitive contributions 
between the active $\gamma$-band and the passive ($\alpha, \beta$)-bands.
Thus, the superconductivity which is mainly caused by the quasi-two-dimensional $\gamma$-band 
may be accompanied by a tiny spin-orbit coupling $\eta \sim 0.001$, 
consistent with the NMR data. 

A small spin-orbit coupling is also indicated by the 
observation of the half-quantum vortex~\cite{Jang}. The half-quantum vortex is formed by the $\pi$-rotation 
of $d$-vector around the vortex core~\cite{Volovik-Mineev}. 
This intriguing topological defect is unstable unless the spin-orbit coupling is small~\cite{Kee-Sigrist}. 
Indeed, theoretical studies of the half-quantum vortex in Sr$_2$RuO$_4$ have assumed 
a small spin-orbit coupling~\cite{Kee-Sigrist,Chung2007,Vakaryuk-Leggett}. 
Such a small spin-orbit coupling is compatible with the superconductivity in the $\gamma$-band, but 
incompatible with the quasi-one-dimensional superconductivity driven by the ($\alpha, \beta$)-bands. 
A moderate spin-orbit coupling $\eta \sim 0.1$ is expected in the later (see Table~II). 
Thus, not only the thermodynamic and transport properties~\cite{Maeno2012,Nomura2002,Nomura2005} 
but also the features of the spin-orbit coupling indicate the superconductivity mainly caused by 
the quasi-two-dimensional $\gamma$-band.

\subsection{Colossal effect of broken inversion symmetry}

Our discussions in this section have been based on the inversion symmetry of the crystal structure, as we 
mentioned in \S3.2. 
When the inversion symmetry is broken, for instance near the surface, the spin-orbit coupling in Cooper 
pairs dramatically changes. 

A simple way to describe the spin-orbit coupling in non-centrosymmetric systems 
is to adopt the antisymmetric spin-orbit coupling (such as the Rashba spin-orbit coupling), 
$H_{\rm ASOC} = \alpha \sum_{\k} \g(\k) \SS(\k)$~\cite{Springer}. 
The antisymmetric spin-orbit coupling gives rise to a large anisotropy in spin-triplet Cooper pairs, $\eta = O(1)$, 
when $|\alpha| > T_{\rm c}$, and it stabilizes the pairing state with $d$-vector parallel to the $g$-vector, 
that is, $\dd(\k) \parallel \g(\k)$~\cite{Frigeri2004}. 

From the microscopic point of view, the antisymmetric spin-orbit coupling arises from the combination of 
the LS coupling and the parity mixing in local electron orbitals.~\cite{YanaseCePt3Si} 
The latter is taken into account in the three-orbital Hubbard model by adding the parity mixing 
term,~\cite{Yanase2013} 
\begin{eqnarray}
&& \hspace*{-14mm} 
\label{H_odd}
H_{\rm odd} = \sum_{\k s} 
[V_{\rm x}(\k) c_{\k 1 s}^{\dag} c_{\k 3 s} 
+V_{\rm y}(\k) c_{\k 2 s}^{\dag} c_{\k 3 s} 
+ {\rm h.c.}].  
\end{eqnarray}
For the extended model $H' = H + H_{\rm odd}$ 
coefficients $A_{nm}$ ($n \ge 1$) in Eq.~(\ref{LS_expansion}) are finite, and therefore, 
the perturbation expansion with respect to the LS coupling is unreliable. 
The non-perturbative calculation shows that a small parity mixing due to the broken inversion symmetry 
stabilizes the $A_{2u}$ state~\cite{Yanase2013}. 
Thus, the spin-triplet SC state is sensitive to the broken inversion symmetry. 

The randomness yielding the locally non-centrosymmetric structure also remarkably affects the spin-triplet 
SC state. For instance, we investigated the roles of the random Rashba spin-orbit coupling induced by 
stacking faults, and found that a small mean square value of Rashba spin-orbit coupling, $\bar{\alpha} =2$ K, stabilizes 
the $A_{2u}$ state~\cite{Yanase2010}. Such SC state may appear in the eutectic crystal 
Sr$_2$RuO$_4$/Sr$_3$Ru$_2$O$_7$ which are indeed influenced by stacking faults~\cite{Kittaka2008}.

\section{Superconducting phases in Sr$_2$RuO$_4$ for $H \parallel [001]$}

 The spin-triplet Cooper pairs in Sr$_2$RuO$_4$ seem to be affected by a small but finite 
spin-orbit coupling $\eta = 0.001 \sim 0.01$, as indicated by both theoretical estimations 
and experimental data (see \S3). Such a small spin-orbit coupling allows the multiple SC 
transitions to occur. In the following part, we theoretically demonstrate multiple SC phases 
in the magnetic field. 
We consider the magnetic field along the crystallographic {\it c}-axis in this section, 
and study the SC state in the magnetic field along the {\it ab}-plane in \S5. 

We assume that the spin-orbit coupling in Cooper pairs is so small that 
a moderate magnetic field below $H_{\rm c2}$ suppresses the $d$-vector parallel to the magnetic field 
through the paramagnetic depairing effect. 
This is likely the case of Sr$_2$RuO$_4$ as we discussed in \S3.4. When the magnetic field 
is parallel to the {\it c}-axis, the chiral state $\dd = (p_{x} \pm i p_{y}) \hat{z} $ ($E_u$ state) 
is destabilized, and other two spin components $d_{x}$ and $d_{y}$ may appear. 
We describe these order parameters in the spin basis 
$\Delta_{\uparrow\uparrow}(\r,\k)$ and $\Delta_{\downarrow\downarrow}(\r,\k)$ instead of the $d$-vector form. 
The quasi-classical form is used for the study of spatially inhomogeneous SC state 
(vortex state). 
Each spin component is divided into the two orbital components, as 
$\Delta_{\sigma\sigma}(\r,\k)=\Delta_{\sigma\sigma,{\rm x}}(\r)\phi_{\rm x}(\k)
+\Delta_{\sigma\sigma,{\rm y}}(\r)\phi_{\rm y}(\k)$, where $\phi_{\rm x}(\k)$ and 
$\phi_{\rm y}(\k)$ stand for pairing functions with the $p_{x}$- and $p_{y}$-wave symmetry, respectively. 
In this way, the SC state is described by the $2 \times 2 = 4$ component order parameters, 
$\left( \Delta_{\uparrow\uparrow,{\rm x}}(\r),  \Delta_{\uparrow\uparrow,{\rm x}}(\r),  
\Delta_{\downarrow\downarrow,{\rm x}}(\r),  \Delta_{\downarrow\downarrow,{\rm y}}(\r) \right)$.

SC state is investigated on the basis of the following GL model~\cite{Takamatsu-Yanase}, 
\begin{equation}
f=\sum_{\sigma=\uparrow,\downarrow} f^{0}_{\sigma} +f^{\rm SO1} + f^{\rm SO2},
\label{GL_c}
\end{equation}
where the first term is the ordinary part of the GL free energy density for the two orbital component 
($p_{x}, p_{y}$)-wave superconductors in the tetragonal lattice~\cite{Agterberg,Agterberg2}, 
\begin{align}
f^{0}_{\sigma}=&\alpha_0 (T -T_{\rm c}^{0}) \bigl(|\Delta_{\sigma\sigma,{\rm x}}|^2+|\Delta_{\sigma\sigma,{\rm y}}|^2\bigr)
\nonumber \\
&
+\beta_1\bigl(|\Delta_{\sigma\sigma,{\rm x}}|^2+|\Delta_{\sigma\sigma,{\rm y}}|^2\bigr)^2/2 \nonumber\\
&+\beta_2(\Delta_{\sigma\sigma,{\rm x}}\Delta^*_{\sigma\sigma,{\rm y}}-{\rm c.c.})^2/2
+\beta_3|\Delta_{\sigma\sigma,{\rm x}}|^2|\Delta_{\sigma\sigma,{\rm y}}|^2 \nonumber\\
&+\xi^2_1\bigl[|D_{\rm x}\Delta_{\sigma\sigma,{\rm x}}|^2
+|D_{\rm y}\Delta_{\sigma\sigma,{\rm y}}|^2\bigr] \nonumber\\
&+\xi^2_2\bigl[|D_{\rm x}\Delta_{\sigma\sigma,{\rm y}}|^2
+|D_{\rm y}\Delta_{\sigma\sigma,{\rm x}}|^2\bigr] \nonumber\\
&+\xi^2_3\Bigl\{\bigl[(D_{\rm x}\Delta_{\sigma\sigma,{\rm x}})(D_{\rm y}\Delta_{\sigma\sigma,{\rm y}})^*
+{\rm c.c.}\bigr] \nonumber\\
& \hspace{10mm} +\bigl[(D_{\rm x}\Delta_{\sigma\sigma,{\rm y}})(D_{\rm y}\Delta_{\sigma\sigma,{\rm x}})^*
+{\rm c.c.}\bigr]\Bigr\}.
\label{eq:f0}
\end{align}
We adopt the conventional notation for the covariant derivative 
$D_{j}=\nabla_{j}+(2\pi{\rm i}/\varPhi_{0})A_{j}$ and $\varPhi_{0}=hc/2|e|$. 
Other notations have been explained in Ref.~\citen{Takamatsu-Yanase}. 
We omitted the label $\r$ to simplify the description of Eq.~(\ref{eq:f0}).

As we have discussed, the spin-orbit coupling plays a crucial role in the spin-triplet superconductors 
in the tetragonal lattice even if it is small. 
Our GL model takes into account two spin-orbit coupling terms,  
\begin{align}
& f^{\rm SO1} = \epsilon \sum_{\sigma} 
\sigma({\rm i}\Delta_{\sigma\sigma,{\rm x}}\Delta^*_{\sigma\sigma,{\rm y}}+{\rm c.c.}),
\label{eq:fso1}
\\
& f^{\rm SO2}=\delta\bigl[(\Delta_{\uparrow\uparrow,{\rm x}}\Delta^*_{\downarrow\downarrow,{\rm x}}
-\Delta_{\uparrow\uparrow,{\rm y}}\Delta^*_{\downarrow\downarrow,{\rm y}})+{\rm c.c.}\bigr].
\label{eq:fso2}
\end{align}
The coupling constants $\epsilon$ and $\delta$ are related with the spin-orbit coupling in 
Cooper pairs as $2 \epsilon = \eta^{A_{1u} B_{1u}} = \eta^{A_{2u} B_{2u}}$ and 
$2 \delta = \eta^{A_{1u} A_{2u}} = \eta^{B_{1u} B_{2u}}$.  
According to the selection rules shown in \S3.3, the $f^{\rm SO1}$ term is given by 
the quasi-one-dimensional ($\alpha,\beta$)-bands. The leading order term has been obtained as $\epsilon = 
O\left(\lambda/E_{\rm F} \cdot t_5 /E_{\rm F} \cdot |\Delta_{\alpha\beta}/\Delta_{\gamma}|^2 \right)$~\cite{Yanase2003}, 
where $\Delta_{\alpha\beta}$ and $\Delta_{\gamma}$ are the magnitude of SC gap in the ($\alpha,\beta$)- and 
$\gamma$-bands, respectively. For $|\Delta_{\alpha\beta}/\Delta_{\gamma}| \sim 0.3 $, we find 
$|\epsilon| = 0.001 \sim 0.01$. 
On the other hand, the $f^{\rm SO2}$ term originates from the coupling between the $\gamma$-band and 
($\alpha,\beta$)-bands. The magnitude has been numerically estimated to be $|\delta| = 0.001 \sim 0.01$ in Fig.~3(b). 
Thus, the magnitudes of two spin-orbit couplings are in the same order. 

We determine the pairing state for temperatures $T$ and magnetic fields $\vec{H} =H \hat{c}$ 
by minimizing the GL free energy with use of the variational method. 
We rewrite the order parameters using the chirality basis, as 
$\Delta_{\sigma\sigma,1} \equiv (\Delta_{\sigma\sigma,{\rm x}}-{\rm i}\Delta_{\sigma\sigma,{\rm y}})/\sqrt{2}$
and
$\Delta_{\sigma\sigma,2} \equiv (\Delta_{\sigma\sigma,{\rm x}}+{\rm i}\Delta_{\sigma\sigma,{\rm y}})/\sqrt{2}$. 
They are assumed to be a linear combination of the basis functions, 
\begin{align}
& \hspace{-3mm}
\left(
\begin{array}{c}
\Delta_{\uparrow\uparrow,1}(\r) \\
\Delta_{\uparrow\uparrow,2}(\r)
\end{array}
\right)
=C_{1}
\left(
\begin{array}{c}
\psi_{1+}(\r,{\bm 0}) \\
\psi_{2+}(\r,{\bm 0}) 
\end{array}
\right)
+C_{2}
\left(
\begin{array}{c}
\psi_{1-}(\r,{\bm \delta_2}) \\
\psi_{2-}(\r,{\bm \delta_2}) 
\end{array}
\right), 
\label{eq:trial_fu_1}
\nonumber \\
\\
& \hspace{-3mm}
\left(
\begin{array}{c}
\Delta_{\downarrow\downarrow,1}(\r)\\
\Delta_{\downarrow\downarrow,2}(\r)
\end{array}
\right)
=C_{3}
\left(
\begin{array}{c}
\psi_{1+}(\r,{\bm \delta_3}) \\
\psi_{2+}(\r,{\bm \delta_3}) 
\end{array}
\right)
+C_{4}
\left(
\begin{array}{c}
\psi_{1-}(\r,{\bm \delta_4}) \\
\psi_{2-}(\r,{\bm \delta_4}) 
\end{array}
\right), 
\label{eq:trial_fu_2}
\nonumber \\
\end{align}
where $\delta_2$, $\delta_3$, $\delta_4$ denote the positions of vortex cores 
in each basis function. We optimize the free energy with respect to these vectors.  
The basis functions $\psi_{j\pm}(\r,{\bm \delta})$ are obtained by solving the linearized GL equation, 
\begin{align}
& \frac{2\pi H}{\varPhi_0}
\left(
\begin{array}{cc}
\kappa(1+2\Pi_+\Pi_-) & -\rho_{-}\Pi^2_+-\rho_{+}\Pi^2_- \\
 -\rho_{+}\Pi^2_+-\rho_{-}\Pi^2_-  & \kappa(1+2\Pi_+\Pi_-)
\end{array}
\right)
\left(
\begin{array}{c}
\psi_{1\pm} \\
\psi_{2\pm}
\end{array}
\right)  \nonumber \\
& =2\lambda_{{\rm min}}^{\pm} 
\left(
\begin{array}{c}
\psi_{1\pm} \\
\psi_{2\pm}
\end{array}
\right). 
\label{eq:tenmoto}
\end{align}
With use of the Landau level expansion, the basis functions are described as 
$\psi_{1+}(\r,{\bm \delta}) = \sum_{n\geq 0} a_{4n} \varphi_{4n}(\r,{\bm \delta})$, 
$\psi_{2+}(\r,{\bm \delta}) = \sum_{n\geq 0} a_{4n+2} \varphi_{4n+2}(\r,{\bm \delta})$
and 
$\psi_{1-}(\r,{\bm \delta}) = \sum_{n\geq 0} b_{4n+2} \varphi_{4n+2}(\r,{\bm \delta})$, 
$\psi_{2-}(\r,{\bm \delta}) = \sum_{n\geq 0} b_{4n} \varphi_{4n}(\r,{\bm \delta})$, respectively. 
The $n$-th Landau  level wave functions are denoted as $\varphi_{n}({\bm r},{\bm 0})$ 
and $\varphi_{n}({\bm r},{\bm \delta})=
e^{-{\rm i}\delta_{\rm x}(y-\delta_{\rm y})}\varphi_{n}({\bm r}-{\bm \delta},{\bm 0})$. 
$\lambda_{{\rm min}}^{+}$ ($\lambda_{{\rm min}}^{-}$) is the minimum eigenvalue of Eq.~(\ref{eq:tenmoto}) 
in the positive (negative) chirality channel. 
Variational parameters $(C_1,C_2,C_3,C_4,{\bm \delta_2},{\bm \delta_3},{\bm \delta_4})$ are
optimized to minimize the GL free-energy.

\begin{figure}[htbp]
\begin{center}
\includegraphics[width=8.5cm]{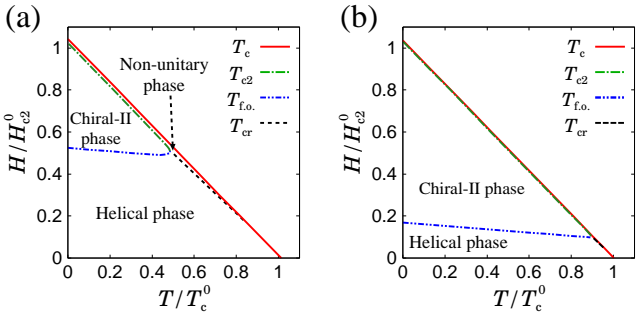}
\end{center}
\caption{(Color online) 
Phase diagram of the four-component GL model [Eq.~(\ref{GL_c})] for temperatures $T$ and 
magnetic fields $H // [001]$~\cite{Takamatsu-Yanase}. 
We assume 
$(\beta_2/\beta_1, \beta_3/\beta_1, \xi_1/\xi, \xi_2/\xi, \xi_3/\xi) = (0.5, 0.5, 1.2, 0.83, 0.3)$.
Spin-orbit couplings are chosen to be (a) $(\epsilon, \delta) = (-0.01, 0.005)$ 
and (b) $(\epsilon, \delta) = (-0.002, 0.001)$. 
The unit of magnetic field is $H^{0}_{\rm c2}=\varPhi_0/2\pi\xi^2$. 
The red solid line shows the SC transition temperature $T_{\rm c}(H)$. 
The green dot-dashed line, blue double-dot-dashed line, and black dashed line 
show the second order transition, first-order transition,  
and crossover in the SC state, respectively. 
}
\label{H_c}
\end{figure}

We show the $H$-$T$ phase diagram for small spin-orbit couplings $(\epsilon,\delta)=(-0.01,0.005)$
and that for tiny spin-orbit couplings $(\epsilon,\delta)=(-0.002,0.001)$ in 
Fig.~\ref{H_c}(a) and Fig.~\ref{H_c}(b), respectively. 
We see that not only the helical state but also the chiral-II state and the non-unitary state 
appear in the phase diagram. The helical state is characterized by the variational parameters 
$|C_1| \sim |C_4|$ and $C_2 = C_3 =0 $, and crossovers to the non-unitary state 
($|C_1| \gg |C_4|$) near $T=T_{\rm c}(H)$. In the high magnetic field region, the chiral-II state 
is stabilized by the coupling of magnetic field and chirality. Then, we obtain the variational parameters, 
$|C_1| \sim |C_3| \gg |C_4| \sim |C_2| > 0$, which is described in the d-vector form as 
$\dd =(p_{x} + i p_{y}) \hat{x}$ or $\dd =(p_{x} + i p_{y}) \hat{y}$. 
This state is distinguished from the chiral state [$\dd =(p_{x} \pm i p_{y}) \hat{z}$] 
because of the difference in the direction of $d$-vector.

Although the $H$-$T$ phase diagram is independent of the sign of spin-orbit couplings, 
the $d$-vector depends on the sign of $\epsilon$ and $\delta$. 
We here assume $\epsilon < 0$ and $\delta > 0$ so that the spin-orbit coupling favors the $A_{2u}$ state 
among the four helical SC states. When we change the sign of $\epsilon$ and $\delta$, the d-vector 
in the helical state changes as summarized in Tables~III. 
The d-vector in the chiral-II and non-unitary states also depend on the sign of $\xi_1 -\xi_2$. 
We assume $\xi_1 -\xi_2 > 0$ in Fig.~4 as usually expected~\cite{Agterberg, Agterberg2}, 
but the perturbation analysis of the three-orbital Hubbard model shows $\xi_1 -\xi_2 < 0$~\cite{Udagawa-Yanase-Ogata}. 
The parameter dependence of the d-vector is summarized in Table~IV.

The $H$-$T$ phase diagram for $H \parallel [001]$ is basically determined by the competition 
between the spin-orbit coupling and the magnetic-field-chirality coupling. 
Generally speaking, the spin-orbit coupling stabilizes 
one of the irreducible representations in Table~I. Indeed, the helical state belonging to the 
$A_{1u}$, $A_{2u}$, $B_{1u}$, or $B_{2u}$ representation is stabilized at low magnetic fields as shown in Fig.~\ref{H_c}. 
On the other hand, the coupling of magnetic field and chirality favors the chiral SC state. 
Although the chirality is canceled out in the helical state, 
the chiral-II state, which belongs to a mixed representation of the $D_{4h}$ point group symmetry, 
has a finite chirality. This is the reason why the chiral-II state is stabilized at high magnetic fields. 
Thus,  the chiral-II phase is stable in the large part of $H$-$T$ phase diagram 
when the spin-orbit coupling is decreased (see Fig.~\ref{H_c}(b)).

Unfortunately, double SC transitions have not been observed in Sr$_2$RuO$_4$ 
in the magnetic field along the {\it c}-axis up to now~\cite{Mackenzie2003,Maeno2012}. 
This experimental status is consistent with our result for tiny spin-orbit couplings (Fig.~\ref{H_c}(b)).  
We again stress that such a tiny spin-orbit coupling is compatible with the 
microscopic estimations based on the three-orbital Hubbard model. 
Then, the helical phase at low magnetic fields may be masked by the Meissner phase, 
which is not negligible in Sr$_2$RuO$_4$ having a moderate Ginzburg parameter 
$\kappa \sim 2.6$~\cite{Maeno2012}. 
If so, the vortex state in the {\it c}-axis magnetic field is the chiral-II phase. 
This phase has an intriguing property, that is, the fractional vortex lattice. 
We showed that the fractional vortices accompanied by the Majorana zero mode~\cite{Ivanov} 
form the lattice owing to the spin-orbit coupling.~\cite{Takamatsu-Yanase} 
This is in sharp contrast to the fact that the fractional vortex is destabilized 
by the spin-orbit coupling in non-chiral spin-triplet superconductors~\cite{Chung}. 
In other words, the chiral spin-triplet superconductor will be a good platform of the fractional vortex. 
Although the vortex lattice at low magnetic fields has been clarified by the small angle neutron 
scattering measurement~\cite{Riseman}, future experimental searches for the exotic vortex lattice structure 
at high magnetic fields are desired. 
Another intriguing feature is its topologically non-trivial property of the chiral II state. 
Recently, the chiral-II state has been identified to be a topological crystalline superconductor~\cite{Ueno}.

\section{Superconducting phases in Sr$_2$RuO$_4$ for $H \parallel [100]$}

Now we turn to the SC phases in the magnetic field along the crystallographic {\it a}- or {\it b}-axis. 
The $E_{u}$ state is robust against the paramagnetic depairing effect for this field direction. 
Since the $E_{u}$ state has two orbital components, 
the SC double transition occurs; the chiral state [$\dd = (\px \pm i \py) \hat{z}$] changes 
to the non-chiral state [$\dd = \px \hat{z}$ or $\dd = \py \hat{z}$] at a moderate magnetic field. 
This chiral to non-chiral transition was predicted by Agterberg using the GL theory~\cite{Agterberg}.

\begin{figure}[htbp]
\begin{center}
\includegraphics[width=7.5cm]{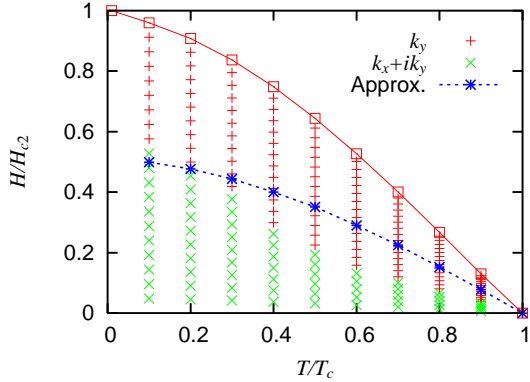}
\end{center}
\caption{(Color online) 
Phase diagram of two component $(\px, \py)$-wave superconductor obtained by 
the quasi-classical Eilenberger equation~\cite{Udagawa_thesis}. 
Red marks $+$ show the high field state $\dd = \py \hat{z}$, and green marks $\times$ show 
the low field state $\dd = (\delta \px + i \py) \hat{z}$. 
Dashed line shows the phase boundary determined by the approximate analytical solution of the 
Eilenberger equation (Pesch's approximation). 
}
\label{H_ab}
\end{figure}

We investigated the 
chiral to non-chiral transition using the quantitatively appropriate calculation 
based on the quasi-classical theory~\cite{Udagawa_thesis}. 
We numerically solve the Eilenberger equation for two component $(\px, \py)$-wave superconductors
with use of the Riccati equation and self-consistently determine the order parameter $\hat{\Delta}(\r,\k_{\rm F}) 
= \hat{\sigma}_{x} \left(\Delta_{x}(\r) \phi_{x}(\k_{\rm F}) + \Delta_{y}(\r) \phi_{y}(\k_{\rm F})\right)$
and the vector potential $A(\r)$. 
The chiral to non-chiral transition occurs as predicted by Agterberg (see Figure~\ref{H_ab}). 
We find that the $\px$-wave component $\Delta_{x}(\r)$ vanishes at high fields  
when we choose $\phi_{x}(\k_{\rm F})$ and $\phi_{y}(\k_{\rm F})$ in accordance with the analysis of 
three-orbital Hubbard model~\cite{Udagawa-Yanase-Ogata}. 
Thus, the non-chiral state ($\dd = \py \hat{z}$) is stabilized in the high magnetic field region. 
The chiral state $\dd = (\delta \px + i \py) \hat{z}$ ($0 \le \delta \le 1$) in the low magnetic field region 
adiabatically changes to the zero-field state $\dd = (\px + i \py) \hat{z}$.

Let us discuss the experimental data of Sr$_2$RuO$_4$. 
Figure~\ref{H_ab} shows that the chiral to non-chiral transition occurs around $H \sim 0.6 H_{\rm c2}$ 
at low temperatures~\cite{Comment1}. This is in agreement with the experimental observation of the 
magnetization kink at $H = 8 \sim 9$ kOe ($\sim 0.6 H_{\rm c2}$) for $T < 0.6$ K~\cite{Tenya}. 
Thus, the magnetization kink may a fingerprint of the double SC transition. 
On the other hand, specific heat measurements have not observed any indication for 
the double SC transition at moderate magnetic fields~\cite{Mackenzie2003,Maeno2012}. 
According to the quasi-classical theory, 
the specific heat jump is too small to be observed at low temperatures ($T < 0.5 T_{\rm c}$)~\cite{Udagawa_thesis}. 
However, our calculation showed a sizable jump of the specific heat around $T=T_{\rm c}$, which has not been 
observed in experiments. 
This inconsistency may imply that the zero-field state is {\it not} the chiral state ($E_{u}$ state). 
When the zero-field state is a helical state ($A_{1u}$, $A_{2u}$, $B_{1u}$, or $B_{2u}$ state), 
the chiral to non-chiral transition does not occur. 
However, this is not a strong evidence against the $E_{u}$ state, as the B-C phase transition of UPt$_3$ 
was missed in the specific heat measurements~\cite{Joynt}. Recent calculation based on the quasi-classical theory 
also pointed out that the signature of double SC transition disappears when the 
magnetic field is slightly tilted less than $1^\circ$ from the {\it ab}-plane~\cite{Ishihara-Ichioka-Machida}.

\begin{figure}[htbp]
\begin{center}
\includegraphics[width=8.8cm]{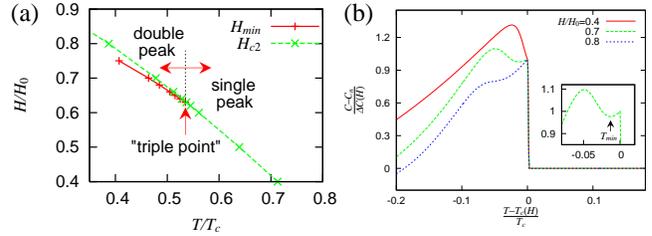}
\end{center}
\caption{(Color online) 
(a) Phase diagram near the upper critical field for $H \parallel [100]$~\cite{Udagawa_thesis}. 
The green dashed line shows the upper critical field, and the red solid line shows the 
crossover from the non-unitary state [$\dd = \py (\hat{z} - i \hat{y})$] to the unitary state [$\dd = \py \hat{z}$]. 
(b) The specific heat shows a double peak (single peak) near \Tc in the high (low) magnetic field region. 
We obtained these results on the basis of the quasi-classical theory of two component 
order parameters, $\py \hat{x}$ and $\py \hat{z}$. The Pesch's approximation was used to solve Elenberger equation. 
}
\label{H_ab_h}
\end{figure}

Finally, we briefly discuss another phase transition near the upper critical field~\cite{Udagawa-Yanase-Ogata}. 
That is  the unitary to non-unitary transition from $\dd = \py \hat{z}$ to 
$\dd = \py (\hat{z} -i \hat{y})$~\cite{Comment2}. 
Strictly speaking, this transition is a crossover in the presence of the spin-orbit coupling. 
However, the specific heat shows a peak, 
when the spin-orbit coupling is tiny $\eta \sim 0.001$~\cite{Udagawa-Yanase-Ogata}. 
Indeed, both GL theory~\cite{Udagawa-Yanase-Ogata} and quasi-classical theory~\cite{Udagawa_thesis} (see Fig.~6) 
reproduces the features of specific heat data indicating the double SC transition~\cite{Deguchi}. 
However, recently observed first order transition at $H=H_{\rm c2}$~\cite{Yonezawa} 
is not reproduced.  Because any weak coupling theory for the spin-triplet superconductivity 
is not compatible with the first order SC transition with a sizable latent heat, it is desired to examine the  
strong coupling effect, which sometimes changes the thermodynamic properties of superconductors. 
In order to explain the first order SC transition, the spin-singlet superconductivity has been considered for  
Sr$_2$RuO$_4$~\cite{Machida-Ichioka} in spite of several contradictory experiments~\cite{Maeno2012}.

\section{Summary and Discussion}

In this article we reviewed the spin-orbit coupling in spin-triplet Cooper pairs and multiple SC phases 
in Sr$_2$RuO$_4$.  
We demonstrated that the spin-orbit coupling arises from the LS coupling of electrons. 
Interestingly, not only the magnitude but also the roles of the spin-orbit coupling are determined 
by the {\it selection rules} which originate from the symmetries of the local electron orbital, 
crystal structure, and superconductivity. 
Therefore, the anisotropy and the easy axis of the $d$-vector are determined 
without relying on the microscopic calculation. 

Based on the selection rules and microscopic analysis of the three-orbital Hubbard model for Sr$_2$RuO$_4$ 
we found that the chiral SC state ($E_{u}$ state) can be stabilized when the quasi-two-dimensional 
$\gamma$-band is mainly superconducting. On the other hand, 
one of the helical states ($A_{1u}$, $A_{2u}$, $B_{1u}$, or $B_{2u}$ state) is stable 
when the quasi-one-dimensional ($\alpha$, $\beta$)-bands are responsible for the superconductivity. 
The spin-orbit coupling of Cooper pairs is tiny in the former case, although the LS coupling is much larger 
than the energy scale of the superconductivity. On the other hand, a moderate spin-orbit coupling appears 
in the latter case. 

It has been claimed that 
the superconductivity in Sr$_2$RuO$_4$ is likely caused by the $\gamma$-band, 
as it is indicated by the thermodynamic and transport properties.~\cite{Mackenzie2003,Maeno2012} 
The features of the spin-orbit coupling also point to this case according to the 
comparison with NMR measurements~\cite{Ishida,Murakawa2004,Murakawa2007} and the observation of 
the half-quantum vortex~\cite{Jang} and TRSB~\cite{Luke1998,Xia2006}. 
However, some controversial data remains to be resolved. 
For instance, the chiral edge mode has not been detected~\cite{Kallin2012}, 
and the O-site NQR measurement~\cite{Mukuda} implies the helical spin-triplet pairing state~\cite{Miyake_NQR}. 
These seemingly controversial data may be understood by considering the small spin-orbit coupling in 
Cooper pairs which allows various textures near the edges, domain walls, and impurities. 

When the spin-orbit coupling in spin-triplet Cooper pairs is small as estimated by the microscopic calculation, 
multiple SC phase transitions occur in the magnetic field. 
We elucidated the $H$-$T$ phase diagram for both field directions $H \parallel [001]$ and 
$H \parallel [100]$ on the basis of the GL theory and quasi-classical theory taking account of a 
small spin-orbit coupling. For $H \parallel [100]$, the unitary to non-unitary transition occurs 
near the upper critical field, and chiral to non-chiral transition occurs at moderate magnetic fields. 
For $H \parallel [001]$, the high field SC phase is the chiral-II state 
[$\dd = (\px + i \py) \hat{x}$ or $\dd = (\px + i \py) \hat{y}$] which accompanies the fractional vortex lattice. 
These SC phases are fingerprint of the small spin-orbit coupling. 
We discussed experimental indications for the multiple SC phases in Sr$_2$RuO$_4$, but 
convincing evidence for them is still on the hunt. 
As the observation of multiple phases in UPt$_3$ has been an convincing evidence 
for the spin-triplet superconductivity~\cite{Sauls,Joynt},  
it is desirable to elucidate whether multiple SC phases appear in Sr$_2$RuO$_4$ or not. 
Our studies provide a basis for the future experimental test.

\section*{Acknowledgements}
A part of this work was carried out in collaboration with M. Mochizuki and M. Ogata. 
The authors are grateful to D. F. Agterberg, K. Deguchi, K. Ishida, S. Kittaka, Y. Maeno, K. Miyake, 
T. Nomura, M. Sigrist, K. Tenya, K. Yamada, and S. Yonezawa for fruitful discussions. 
This work was supported by KAKENHI (Nos. 24740221, 24740230, and 25103711).
Part of numerical computation in this work was carried out 
at the Yukawa Institute Computer Facility.



\newpage

\begin{table*}[htbp]
\begin{center}
{\renewcommand\arraystretch{1.4}
\begin{tabular}[htb]{c||c|c|c|c|c} 
Crystal symmetry
& \multicolumn{2}{c}{Tetragonal } 
& \multicolumn{3}{|c}{Hexagonal } 
\\ \hline  
Local electron orbital
& d$_{\rm xy}$ & d$_{\rm yz}$, d$_{\rm zx}$ & 
A$_{\rm 1g}$ & 
\multicolumn{2}{|c}{E$_{\rm g}$} 
\\ \hline 
Orbital symmetry of SC 
& \multicolumn{2}{c|}{P-wave} & P- or F-wave & P-wave & F-wave 
\\ \hline \hline 
Easy axis of $d$-vector 
& $\left( \dd \parallel c \right)$ & $ \dd \parallel ab$ & both & $ \dd \parallel ab$ & both 
\\ \hline
Anisotropy $[ \eta = 1 - T_{\rm c}^{\Gamma_1}/T_{\rm c}^{\Gamma_0} ]$
& $O\left(\lambda^{2}/E_{\rm F}^{2} \right)$ & $O\left(\lambda/E_{\rm F} \right)$ & 
$O\left(\lambda^{2}/E_{\rm F}^{2} \right)$ &
$O\left(\lambda/E_{\rm F} \right)$ & $O\left(\lambda^{2}/E_{\rm F}^{2} \right)$ 
\\ 
\end{tabular}
}
\caption{
Table of spin-triplet superconductivity in $t_{\rm 2g}$ electron systems~\cite{Yanase2003, Yanase2005}. 
We show the selection rules which determine the easy axis and the anisotropy of spin-triplet Cooper pairs 
on the basis of the symmetries of crystal, local electron orbital, and Cooper pairs' orbital.
Although the easy axis is not determined by the selection rule for the d$_{\rm xy}$-orbital in the tetragonal lattice, 
we show the result obtained by the perturbation theory assuming the parameters for \SRO (Fig.~3(a)). 
}
\label{table2}
\end{center}
\end{table*}

\begin{table*}[htbp]
\begin{center}
{\renewcommand\arraystretch{1.2}
\begin{tabular}{c|c|c|c}
       \multicolumn{2}{c|}{$\epsilon>0$}&\multicolumn{2}{c}{$\epsilon<0$}\\ \hline
       $\delta>0$       & $\delta<0$   &   $\delta>0$      & $\delta<0$ \\ \hline\hline
$p_{\rm x}{\hat x}+p_{\rm y}{\hat y}$  & $p_{\rm y}{\hat x}-p_{\rm x}{\hat y}$ 
&$p_{\rm x}{\hat x}-p_{\rm y}{\hat y}$ & $p_{\rm y}{\hat x}+p_{\rm x}{\hat y}$ \\ 
\end{tabular}
}
\end{center}
\caption{
The d-vector in the helical state (last row) when we assume the sign of 
couplings $\epsilon$ and $\delta$ as in the first and second row, respectively.}
\label{table3}
\end{table*}

\begin{table*}[htbp]
\begin{center}
{\renewcommand\arraystretch{1.3}
\begin{tabular}{c||c|c|c|c}
                  &\multicolumn{2}{c|}{$\xi_1>\xi_2$}  &  \multicolumn{2}{c}{$\xi_1<\xi_2$} \\ \hline
                  &  $\epsilon>0$       & $\epsilon<0$    &   $\epsilon>0$   & $\epsilon<0$ \\ 
\hline\hline 
Chiral II state   &\multicolumn{2}{c|}{$(p_{\rm x}+{\rm i}p_{\rm y}){\hat x},~(p_{\rm x}+{\rm i}p_{\rm y}){\hat y}$}
                  &\multicolumn{2}{c}{$(p_{\rm x}-{\rm i}p_{\rm y}){\hat x},~(p_{\rm x}-{\rm i}p_{\rm y}){\hat y}$}\\
\hline
Non-unitary state &$({\hat x}-{\rm i}{\hat y}) (p_{\rm x}+{\rm i}p_{\rm y})$&
                   $({\hat x}+{\rm i}{\hat y}) (p_{\rm x}+{\rm i}p_{\rm y})$&
                   $({\hat x}+{\rm i}{\hat y}) (p_{\rm x}-{\rm i}p_{\rm y})$&
                   $({\hat x}-{\rm i}{\hat y}) (p_{\rm x}-{\rm i}p_{\rm y})$ \\ 
\end{tabular}
}
\end{center}
\caption{
The d-vector in the chiral II state (third row) and in the non-unitary state (fourth row). 
The chirality in these states depends on the magnitude relation between $\xi_1$ and $\xi_2$ (first row). 
The spin component in the non-unitary state also depends on the sign of $\epsilon$ (second row).
The two-fold degeneracy between $\dd \parallel \hat{x}$ and $\dd \parallel \hat{y}$ is not lifted 
in the chiral-II state. 
} 
\label{table4}
\end{table*}




\begin{thebibliography}{10}



\bibitem{Maeno1994}
Y. Maeno, H. Hashimoto, K. Yoshida, S. Nishizaki, T. Fujita, J.~G. Bednorz, and
F. Lichtenberg:  Nature {\bf 372} (1994) 532.

\bibitem{Mackenzie2003}
A.~P. Mackenzie and Y. Maeno:  Rev. Mod. Phys. {\bf 75} (2003) 657. 

\bibitem{Maeno2012}
Y. Maeno, S. Kittaka, T. Nomura, S. Yonezawa, and K. Ishida: 
J. Phys. Soc. Jpn. {\bf 81} (2012) 011009. 


\bibitem{Leggett1975}
A.~J. Leggett:  Rev. Mod. Phys. {\bf 47} (1975) 331.

\bibitem{Sauls}
J. A. Sauls: Adv. Phys. {\bf 43} (1994) 153. 

\bibitem{Joynt}
R. Joynt and L. Taillefer: Rev. Mod. Phys. {\bf 74} (2002) 235. 

\bibitem{Tou}
H. Tou, Y. Kitaoka, K. Asayama, N. Kimura, Y. Onuki, E. Yamamoto, and 
K. Maezawa: Phys. Rev. Lett. {\bf 77} (1996) 1374; 
H. Tou, Y. Kitaoka, K. Ishida, K. Asayama, N. Kimura, Y. Onuki, E. Yamamoto, 
Y. Haga, and K. Maezawa: Phys. Rev. Lett. {\bf 80} (1998) 3129. 

\bibitem{Aoki}
For a review, D. Aoki and J. Flouquet: 
J. Phys. Soc. Jpn. {\bf 81} (2012) 011003. 


\bibitem{Sigrist-Ueda}
M. Sigrist and K. Ueda:  Rev. Mod. Phys. {\bf 63} (1991) 239.


\bibitem{Luke1998}
G. M. Luke, Y. Fudamoto, K. M. Kojima, M. I. Larkin, J. Merrin, B. Nachumi, Y. J. Uemura, 
Y. Maeno, Z. Q. Mao, Y. Mori, H. Nakamura, and M. Sigrist: Nature {\bf 374} (1998) 558. 

\bibitem{Xia2006}
J. Xia, Y. Maeno, P. T. Beyersdorf, M. M. Fejer, and A. Kapitulnik: 
Phys. Rev. Lett. {\bf 97} (2006) 167002. 


\bibitem{Harima}
Y. Yanase and H. Harima: Kotai Butsuri {\bf 46} (2011) 229 (in Japanese). 

\bibitem{Yosida_book}
K. Yosida: {\it ``Theory of Magnetism''} (Springer-Verlag, 1996). 



\bibitem{Yanase2003}
Y. Yanase and M. Ogata:  J. Phys. Soc. Jpn.  {\bf 72} (2003) 673.

\bibitem{Yanase2005}
Y. Yanase, M, Mochizuki and M. Ogata: 
J. Phys. Soc. Jpn. {\bf 74} (2005) 2568. 


\bibitem{Oguchi}
T. Oguchi: Phys. Rev. B {\bf 51} (1995) 1385. 

\bibitem{Singh}
D. J. Singh: Phys. Rev. B {\bf 52} (1995) 1358. 

\bibitem{Yanase_review}
Y. Yanase, T. Jujo, T. Nomura, H. Ikeda, T. Hotta, and K. Yamada: 
Phys. Rep. {\bf 387} (2003) 1. 


\bibitem{Nomura2002-2}
T. Nomura and K. Yamada: J. Phys. Soc. Jpn. {\bf 71} (2002) 1993. 


\bibitem{Hoshihara}
K. Hoshihara and K. Miyake: J. Phys. Soc. Jpn. {\bf 74} (2005) 2679. 


\bibitem{Wang}
Q.-H. Wang, C. Platt, Y. Yang, C. Honerkamp, F. C. Zhang, W. Hanke, 
T. M. Rice, and R. Thomale: Europhys. Lett. {\bf 104} (2013) 17013. 


\bibitem{Takimoto}
T. Takimoto: Phys. Rev. B {\bf 62} (2000) R14641. 


\bibitem{Raghu}
S. Raghu, A. Kapitulnik, and S. A. Kivelson: 
Phys. Rev. Lett. {\bf 105} (2010) 136401. 


\bibitem{Agterberg1997}
D. F. Agterberg, T. M. Rice, and M. Sigrist: 
Phys. Rev. Lett. {\bf 78} (1997) 3374. 


\bibitem{Ng-Sigrist}
K. K. Ng and M. Sigrist: Europhys. Lett. {\bf 49} (2000) 473. 


\bibitem{Yoshioka}
Y. Yoshioka and K. Miyake: J. Phys. Soc. Jpn. {\bf 78} (2009) 074701. 



\bibitem{Kallin2012}
For a review, C. Kallin: Rep. Prog. Phys. {\bf 75} (2012) 042501. 


\bibitem{Ishida}
K. Ishida, H. Mukuda, Y. Kitaoka, K. Asayama, Z. Q. Mao, Y. Mori, 
and Y. Maeno: Nature {\bf 396} (1998) 658. 


\bibitem{Murakawa2004}
H. Murakawa, K. Ishida, K. Kitagawa, Z.~Q. Mao, and Y. Maeno:  Phys. Rev. Lett.
  {\bf 93} (2004) 167004.

\bibitem{Murakawa2007}
H. Murakawa, K. Ishida, K. Kitagawa, H. Ikeda, Z.~Q. Mao, and Y. Maeno:
  J. Phys. Soc. Jpn.  {\bf 76} (2007) 024716. 


\bibitem{Kitagawa2007}
K. Kitagawa, K. Ishida, R. S. Perry, H. Murakawa, K. Yoshimura, and Y. Maeno:
Phys. Rev. B {\bf 75} (2007) 024421. 


\bibitem{Jang}
J. Jang, D. G. Ferguson, V. Vakaryuk, R. Budakian, S. B. Chung, 
P. M. Goldbard, and Y. Maeno: Science {\bf 331} (2011) 186.

\bibitem{Volovik-Mineev}
G. E. Volovik and V. P. Mineev: JETP Lett. {\bf 24} (1976) 561. 


\bibitem{Kee-Sigrist}
H.-Y. Kee and M. Sigrist: arXiv:1307.5859. 


\bibitem{Chung2007}
S. B. Chung, H. Bluhm, E.-A. Kim, Phys. Rev. Lett. {\bf 99} (2007) 197002.

\bibitem{Vakaryuk-Leggett}
V. Vakaryuk and A. J. Leggett, Phys. Rev. Lett. {\bf 103} (2009) 057003.


\bibitem{Nomura2002}
T. Nomura and K. Yamada: J. Phys. Soc. Jpn. {\bf 71} (2002) 404. 

\bibitem{Nomura2005}
T. Nomura: J. Phys. Soc. Jpn. {\bf 74} (2005) 1818.





\bibitem{Springer}
{\it Non-Centrosymmetric Superconductors: Introduction and Overview},
ed. by E. Bauer and M. Sigrist (Springer-Verlag, 2012). 



\bibitem{Frigeri2004}
P.~A. Frigeri, D.~F. Agterberg, A. Koga, and M. Sigrist:  
Phys. Rev. Lett. {\bf 92} (2004) 097001. 

\bibitem{YanaseCePt3Si} 
Y. Yanase and M. Sigrist: 
J. Phys. Soc. Jpn. {\bf 77} (2008) 124711. 

\bibitem{Yanase2013}
Y. Yanase: J. Phys. Soc. Jpn. {\bf 82} (2013) 044711.

\bibitem{Yanase2010}
Y. Yanase: J. Phys. Soc. Jpn. {\bf 79} (2010) 084701. 

\bibitem{Kittaka2008}
S. Kittaka, S. Fusanobori, S. Yonezawa, H. Yaguchi, Y. Maeno, R. Fittipaldi,
and A. Vecchione:  Phys. Rev. B {\bf 77} (2008) 214511. 




\bibitem{Takamatsu-Yanase}
S. Takamatsu and Y. Yanase: J. Phys. Soc. Jpn. {\bf 82} (2013) 063706. 


\bibitem{Agterberg}
D. F. Agterberg: Phys. Rev. Lett. {\bf 80} (1998) 5184. 

\bibitem{Agterberg2}
D. F. Agterberg: Phys. Rev. B {\bf 58} (1998) 14484. 



\bibitem{Udagawa-Yanase-Ogata}
M. Udagawa, Y. Yanase, and M. Ogata: 
J. Phys. Soc. Jpn. {\bf 74} (2005) 2905.


\bibitem{Ivanov}
D. A. Ivanov: Phys. Rev. Lett. {\bf 86} (2001) 268. 

\bibitem{Chung} 
S. B. Chung, D. F. Agterberg, and E.-A. Kim: 
New J. Phys. {\bf 11} (2009) 085004. 


\bibitem{Riseman}
T. M. Riseman, P. G. Kealy, E. M. Forgan, A. P. Mackenzie, L. M. Galvin, 
A. W. Tyler, S. L. Lee, C. Ager, D. McK. Paul, C. M. Aegerter, R. Cubitt, 
Z. Q. Mao, T. Akima, and Y. Maeno: 
Nature {\bf 396} (1998) 19; {\bf 404} (2000) 629. 


\bibitem{Ueno}
Y. Ueno, A. Yamakage, Y. Tanaka, and M. Sato:
Phys. Rev. Lett. {\bf 111} (2013) 087002. 


\bibitem{Udagawa_thesis}
M. Udagawa: Doctor Thesis in University of Tokyo (2007). 


\bibitem{Comment1}
The second order chiral to non-chiral phase transition occurs at moderate magnetic fields 
unless we precisely adjust the parameters so that $\xi_1 \approx \xi_2$~\cite{Kaur}. 

\bibitem{Kaur}
R. P. Kaur, D. F. Agterberg, and H. Kusunose: Phys. Rev. B {\bf 72} (2005) 144528. 

\bibitem{Tenya}
K. Tenya, S. Yasuda, M. Yokoyama, H. Amitsuka, K. Deguchi, and Y. Maeno: 
J. Phys. Soc. Jpn. {\bf 75} (2006) 023702. 


\bibitem{Ishihara-Ichioka-Machida}
M. Ishihara, Y. Amano, M. Ichioka, and K. Machida: Phys. Rev. B {\bf 87} (2013) 224509. 


\bibitem{Comment2}
The unitary to non-unitary transition is similar to the $A_1$-$A_2$ transition 
of superfluid $^{3}$He~\cite{Leggett1975}. 


\bibitem{Deguchi}
K. Deguchi, M. A. Tanatar, Z. Mao, T. Ishiguro, and Y. Maeno: 
J. Phys. Soc. Jpn. {\bf 71} (2002) 2839.

\bibitem{Yonezawa}
S. Yonezawa, T. Kajikawa, and Y. Maeno: 
Phys. Rev. Lett. {\bf 110} (2013) 077003. 

\bibitem{Machida-Ichioka}
K. Machida and M. Ichioka: Phys. Rev. B {\bf 77} (2008) 184515.



\bibitem{Mukuda}
H. Mukuda, K. Ishida, Y. Kitaoka, K. Miyake, Z. Q. Mao, Y. Mori, and Y. Maeno: 
Phys. Rev. B {\bf 65} (2002) 132507. 

\bibitem{Miyake_NQR}
K. Miyake: J. Phys. Soc. Jpn. {\bf 79} (2010) 024714. 



\end{thebibliography}
\end{document}